\documentclass[prd,twocolumn,a4paper,superscriptaddress,nofootinbib]{revtex4}
\usepackage{amsmath,epsfig,natbib,bm,psfrag}

\begin{document}


\newcommand{\Tr}{\text{tr}}
\newcommand{\ra}{\rangle}
\newcommand{\la}{\langle}
\newcommand{\Bt}{\tilde{B}}


\newcommand{\Mpc}{\text{Mpc}}
\newcommand{\half}{{\textstyle \frac{1}{2}}}
\newcommand{\third}{{\textstyle \frac{1}{3}}}
\newcommand{\numfrac}[2]{{\textstyle \frac{#1}{#2}}}
\renewcommand{\d}{\text{d}}
\newcommand{\grad}{\nabla}

\newcommand{\begm}{\begin{pmatrix}}
\newcommand{\enm}{\end{pmatrix}}

\newcommand{\threej}[6]{{\begm #1 & #2 & #3 \\ #4 & #5 & #6 \enm}}
\newcommand{\fsky}{f_{\text{sky}}}

\newcommand{\cla}{\mathcal{A}}
\newcommand{\clb}{\mathcal{B}}
\newcommand{\clc}{\mathcal{C}}
\newcommand{\cle}{\mathcal{E}}
\newcommand{\clf}{\mathcal{F}}
\newcommand{\clg}{\mathcal{G}}
\newcommand{\clh}{\mathcal{H}}
\newcommand{\cli}{\mathcal{I}}
\newcommand{\clj}{\mathcal{J}}
\newcommand{\clk}{\mathcal{K}}
\newcommand{\cll}{\mathcal{L}}
\newcommand{\clm}{\mathcal{M}}
\newcommand{\cln}{\mathcal{N}}
\newcommand{\clo}{\mathcal{O}}
\newcommand{\clp}{\mathcal{P}}
\newcommand{\clq}{\mathcal{Q}}
\newcommand{\clr}{\mathcal{R}}
\newcommand{\cls}{\mathcal{S}}
\newcommand{\clt}{\mathcal{T}}
\newcommand{\clu}{\mathcal{U}}
\newcommand{\clv}{\mathcal{V}}
\newcommand{\clw}{\mathcal{W}}
\newcommand{\clx}{\mathcal{X}}
\newcommand{\cly}{\mathcal{Y}}
\newcommand{\clz}{\mathcal{Z}}
\newcommand{\CMBFAST}{\textsc{cmbfast}}
\newcommand{\CAMB}{\textsc{camb}}
\newcommand{\Omtot}{\Omega_{\mathrm{tot}}}
\newcommand{\Omb}{\Omega_{\mathrm{b}}}
\newcommand{\Omc}{\Omega_{\mathrm{c}}}
\newcommand{\Omm}{\Omega_{\mathrm{m}}}
\newcommand{\omb}{\omega_{\mathrm{b}}}
\newcommand{\omc}{\omega_{\mathrm{c}}}
\newcommand{\omm}{\omega_{\mathrm{m}}}
\newcommand{\Omdm}{\Omega_{\mathrm{DM}}}
\newcommand{\Omnu}{\Omega_{\nu}}

\newcommand{\Oml}{\Omega_\Lambda}
\newcommand{\OmK}{\Omega_K}

\newcommand{\Hunit}{~\text{km}~\text{s}^{-1} \Mpc^{-1}}
\newcommand{\Gyr}{{\rm Gyr}}

\newcommand{\nrun}{n_{\text{run}}}

\newcommand{\lmax}{\ell_{\text{max}}}

\newcommand{\zre}{z_{\text{re}}}
\newcommand{\mpl}{m_{\text{Pl}}}

\newcommand{\vv}{\mathbf{v}}
\newcommand{\vd}{\mathbf{d}}
\newcommand{\vC}{\mathbf{C}}
\newcommand{\vT}{\mathbf{T}}
\newcommand{\eV}{\,\text{eV}}
\newcommand{\vtheta}{\bm{\theta}}
\newcommand{\tT}{\tilde{T}}
\newcommand{\tE}{\tilde{E}}
\newcommand{\tB}{\tilde{B}}

\newcommand{\bld}[1]{\mathrm{#1}}
\newcommand{\mC}{\bm{C}}
\newcommand{\mQ}{\bm{Q}}
\newcommand{\mU}{\bm{U}}
\newcommand{\mN}{\bm{N}}
\newcommand{\mX}{\bm{X}}
\newcommand{\mV}{\bm{V}}
\newcommand{\mP}{\bm{P}}
\newcommand{\mR}{\bm{R}}
\newcommand{\mW}{\bm{W}}
\newcommand{\mD}{\bm{D}}
\newcommand{\mI}{\bm{I}}
\newcommand{\mH}{\bm{H}}
\newcommand{\mM}{\bm{M}}
\newcommand{\mS}{\bm{S}}
\newcommand{\mzero}{\bm{0}}

\newcommand{\btheta}{\bm{\theta}}
\newcommand{\bphi}{\bm{\phi}}

\newcommand{\vA}{\mathbf{A}}
\newcommand{\vAt}{\tilde{\mathbf{A}}}
\newcommand{\ve}{\mathbf{e}}
\newcommand{\vE}{\mathbf{E}}
\newcommand{\vB}{\mathbf{B}}
\newcommand{\vEt}{\tilde{\mathbf{E}}}
\newcommand{\vBt}{\tilde{\mathbf{B}}}
\newcommand{\vEw}{\mathbf{E}_W}
\newcommand{\vBw}{\mathbf{B}_W}
\newcommand{\vx}{\mathbf{x}}
\newcommand{\vX}{\mathbf{X}}
\newcommand{\vXt}{\tilde{\vX}}
\newcommand{\vY}{\mathbf{Y}}
\newcommand{\vBwr}{{\vBw^{(R)}}}
\newcommand{\RW}{{W^{(R)}}}

\newcommand{\mUt}{\tilde{\mU}}
\newcommand{\mVt}{\tilde{\mV}}
\newcommand{\mDt}{\tilde{\mD}}

\newcommand{\Rot}{\begm \mzero &\mI \\ -\mI & \mzero \enm}
\newcommand{\Pt}{\begm \vEt \\ \vBt \enm}


\title{Harmonic E/B decomposition for CMB polarization maps}

\author{Antony Lewis}
 \email{antony@cosmologist.info}
 \affiliation{CITA, 60 St. George St, Toronto M5S 3H8, ON, Canada.}

\begin{abstract}
\vspace{\baselineskip}

The full sky cosmic microwave background polarization field can be
decomposed into `electric' (E) and `magnetic' (B) components that are
signatures of distinct physical processes. 
We give a general construction that achieves 
separation of E and B modes
on arbitrary sections of the sky at the expense of increasing the
noise. 
When E modes are present on all scales the separation
of all of the B signal is no longer possible: there are 
inevitably ambiguous modes that cannot be separated. We discuss the
practicality of performing E/B decomposition on large scales with
realistic non-symmetric sky-cuts, and show that separation on large
scales is possible by retaining only the well supported modes. The
large scale modes potentially contain a great deal of useful
information, and E/B separation at the level of the map is essential
for clean detection of B without confusion from cosmic variance due to
the E signal. We give simple matrix manipulations for creating pure E and B
maps of the large scale signal for general sky cuts. We demonstrate
that the method works well in a realistic case and give estimates of
the performance with data from the Planck satellite.
In the appendix
we discuss the simple analytic case of an azimuthally symmetric cut, and
show that exact E/B separation is possible on an azimuthally symmetric
cut with a finite number of non-intersecting circular cuts around foreground sources.

\end{abstract}


\maketitle


Observations of fluctuations in the cosmic
microwave background (CMB) provide information about
primordial inhomogeneities in the universe. 
One of the most interesting questions 
is whether there was a tensor (gravitational wave) component.
CMB polarization 
measurements offer a probe of this
signal~\cite{Kamionkowski97,Zaldarriaga97,Hu98} via the `magnetic'
component of the polarization map.  This offers the opportunity
to distinguish between different models of single field inflation, which
generically predict a significant amplitude of gravitational waves, and 
other competing models which predict nearly zero amplitudes. 
For example the ekpyrotic scenario generically predicts exponentially small amplitudes
of tensor modes, as do inflation models where density
perturbations originate from fluctuations in a second scalar field.

Polarization of the cosmic microwave sky is produced by electron
scattering, as photons decouple from the primordial plasma and during
reionization.  Gravitational waves produce `magnetic' (B) and `electric'
(E) polarization components at a comparable level
by anisotropic redshifting of the energy of photons. Magnetic
polarization is not produced by linear scalar (density) perturbations, so
detection of a magnetic component would provide strong direct evidence
for the presence of a primordial gravitational wave (tensor)
component. There is a non-linear contribution to the magnetic
signal from gravitational lensing of E polarization, though on the
scales where the tensor mode signal is large the lensing signal is
sufficiently small that it is negligible for observations up to Planck\footnote{\url{http://astro.estec.esa.nl/Planck}} 
sensitivity. However ultimately the lensing signal (and how well it can be
subtracted) may provide a limit on the amplitude of tensor modes that
can be detected~\cite{Knox02,Okamoto03,Hirata:2003ka}. There are also
a variety of other possible sources for tensor and vector modes
generating a B mode
signature~\cite{Seljak:1997ii,Pogosian:2003mz,Lewis:2004ef,Lewis:2004kg},
as well as numerous systematics~\cite{Hu:2002vu}.

Inflationary models generically predict a Gaussian spectrum of linear E and B modes,
in which case it is possible to use the full likelihood function to
constrain the tensor amplitude without separating into E and B modes.
However if there are any departures from
Gaussianity, e.g. due to systematics, foregrounds, magnetic
fields~\cite{Lewis:2004ef},
defects~\cite{Seljak:1997ii,Pogosian:2003mz} or unexpected physics, this could give misleading answers. 
So it is useful to have methods to
separate out pure B modes for robust detections and isolation of
unexpected features.
For small tensor amplitudes any mixture of E and B modes would be dominated by the 
scalar E signal, and a joint analysis for the tensor contribution would be only marginally
more optimal than using only the pure B modes.

On small scales it is possible to provide excellent
separation of the E and B mode power spectra using fast quadratic
estimators~\cite{Chon:2003gx,Crittenden00}. These methods provide estimators which, when averaged over
realizations, give zero if there really is no B signal. However in any
given realization, such as the one we observe, there will in general
be a non-zero B signal. This is a potential obstacle to clean E/B power spectrum separation on
the large scales where cosmic variance is most important.

CMB observations by WMAP have recently indicated a substantial optical depth to
reionization~\cite{Kogut:2003et}. If confirmed this would imply that
detection of magnetic polarization could be achieved by
observation of a relatively small number of high signal to noise B modes on large
scales (corresponding to the horizon size at
reionization).
The large scales are just where cosmic variance is largest, and
the effect of incompleteness of the map (e.g. due to cuts around the
galaxy and foreground sources) makes the E and B modes harder to separate.
There is therefore clear motivation for methods to cleanly separate
the E and B modes on the largest scales, avoiding problems with
 cosmic-variance mixing that arises from the use of quadratic
estimators, and extracting the B modes without assumptions about their
distribution.

In this paper we
discuss various harmonic methods for performing E/B separation,
equivalent to separation at the level of the map. We start in
Section~\ref{EB} with a brief summary of the E/B decomposition. In
Section~\ref{harmonics} we review the tensor harmonics and how the E/B
mixing enters for observations over only part of the sky, and discuss
the theoretical properties of the coupling matrices. In Section~\ref{separation}
we show how to separate the modes for band limited and non-band limited fields with
arbitrary sky cuts, and
show that the method presented in earlier work~\cite{Lewis01} is
optimal for general cuts in the all-scales limit. For non-band limited fields there are 
ambiguous modes than cannot be separated.
We then show how the
large scale reionization magnetic modes can be extracted from a
realistic map by retaining only the well supported modes. The
method is computationally tractable and close to exact, and
should allow robust detection of the large scale magnetic signal with
future CMB polarization observations. In Section~\ref{Planckex} we demonstrate the performance
explicitly for a complicated sky cut geometry, and provide comparative
estimates of the ability of
the Planck satellite to detect tensor modes using different
methods. In the Appendix we review some analytic results for
azimuthal cuts from Ref.~\cite{Lewis01}, show that exact separation is
possible for non-intersecting combinations of azimuthal cuts, and give
a slightly improved matrix method for extracting the pure E and B
modes exactly.

For a method similar to the general method presented here, but working explicitly in pixel space,
see Ref.~\cite{Bunn03}, and other related work in Refs.~\cite{Zaldarriaga01,Crittenden00,Chon:2003gx,Park:2003kp}. Though we only discuss CMB polarization maps
explicitly, the techniques could be applied to other data, for example
shear distributions in weak lensing surveys. Indeed B modes in the lensing of 
pre-reionization gas may provide a good alternative way to detect tensor modes~\cite{Pen:2003yv}. 

\section{E/B Polarization}
\label{EB}
The observable polarization field is described in terms of the two
Stokes' parameters $Q$ and $U$ with respect to a particular choice of
axes about each direction on the sky. We use spherical polar
coordinates, with orthonormal basis vectors $\bm{\sigma}_\theta$ and $\bm{\sigma}_\phi$. The Stokes' parameters define a symmetric and trace-free (STF) rank two linear polarization tensor on
the sphere
\begin{equation}
\clp^{ab} = \frac{1}{2}[Q (\bm{\sigma}_\theta^a \bm{\sigma}_\theta^b
- \bm{\sigma}_\phi^a \bm{\sigma}_\phi^b) 
 - U (\bm{\sigma}_\theta^a \bm{\sigma}_\phi^b
+ \bm{\sigma}_\phi^a \bm{\sigma}_\theta^b)]. 
\end{equation}
A two dimensional STF tensor can be written as a sum of
`gradient' and `curl' parts
\begin{equation}
\clp_{ab} = \nabla_{\la a}\nabla_{b\ra}P_E -
\epsilon^c{}_{(a}\nabla_{b)}\nabla_c P_B.
\end{equation}
where $\nabla$ is the covariant derivative on the sphere,
angle brackets denote the STF part on the enclosed indices, and
round brackets denote symmetrization.
The underlying scalar fields $P_E$ and $P_B$ describe
electric and magnetic polarization respectively and are clearly
non-local functions of the Stokes' parameters. One
can define scalar quantities which are local in the polarization by
taking two covariant derivatives to form $\nabla^a\nabla^b \clp_{ab}=
(\nabla^2+2)\nabla^2 P_E$ and
$\epsilon^{b}{}_c\nabla^c\nabla^a \clp_{ab} = (\nabla^2+2)\nabla^2 P_B$
which depend only on the electric and
magnetic polarization respectively. 
For band limited data one could consider taking these derivatives
of the data to extract the E and B components. 
The problem is that in the
neighborhood of boundaries it becomes harder to measure the second
derivatives, so the noise properties are not straightforward: 
taking the derivatives is effectively increasing the noise near the
boundaries relative to the rest of the map. 

Rather than taking second derivatives it it useful to work with
integrals over the surface. As an example we focus here on the B
polarization, since this is of most interest. We define the surface integral
\begin{equation}
B_W \equiv - 2\int_S \text{d}S \,W \epsilon^{b}{}_c\nabla^c\nabla^a
\clp_{ab},
\end{equation}
where $W$ is a real window function defined over some patch $S$
of the observed portion of the sky. The factor of minus two is included to
make our definition equivalent to that in
Ref.~\cite{Lewis01}. Integrating by parts we have
\begin{eqnarray}
B_W &=& \sqrt{2}\int_S \text{d}S\,   W_B^{ab}{}^* \clp_{ab} \nonumber\\
&-&
2\oint_{\partial S}
\text{d}l^a \left(\epsilon^b{}_a W \nabla^c \clp_{cb} -
\epsilon^b{}_c\nabla^c W \clp_{ab}\right),
\end{eqnarray}
where $W_{B\,ab}\equiv \sqrt{2}\epsilon^c{}_{(a} \nabla_{ b)} \nabla_{c} W$ is
an STF tensor window function. Thus extraction of the B polarization
amounts to measuring a well defined surface integral, \emph{and two
  line integrals}. It is these line integrals that encode the
troublesome aspect of the E/B decomposition.

\section{E/B Harmonics}
\label{harmonics}

Since the E/B decomposition is inherently non-local, it is quite natural to work
in harmonic space. The polarization tensor $\clp_{ab}$ can be expanded
over the whole sky in terms of STF tensor harmonics
\begin{equation}
\clp_{ab} = \frac{1}{\sqrt{2}} \sum_{lm} \left( E_{lm}\, Y_{(lm)ab}^G +
B_{lm}\, Y_{(lm)ab}^C\right),
\end{equation}
where $Y_{(lm)ab}^G$ and $ Y_{(lm)ab}^C$ are the gradient and curl tensor
harmonics of opposite parities defined in Ref.~\cite{Kamionkowski97}.
From the orthogonality of the spherical harmonics over the full sphere
it follows that
\begin{eqnarray}
B_{lm} = \sqrt{2}\int_{4\pi} \text{d}S\, Y_{(lm)}^{C\,ab*} \clp_{ab}.
\label{eq:EBlm}
\end{eqnarray}
In a rotationally-invariant ensemble, the expectation values of the harmonic
coefficients define the power spectrum:
\begin{equation}
\la B_{(lm)'}^\ast B_{lm} \ra = \delta_{l'l}\delta_{m'm} C_l^{BB}.
\end{equation}

When we have data over a section of the sphere, the observed data can
be encoded in a set of pseudo-multipoles $\tilde{E}_{lm}$ and
$\tilde{B}_{lm}$ obtained by including a window function $W$ in the
integral of Eq.~\eqref{eq:EBlm}:
\begin{eqnarray}
\tilde{B}_{lm} = \sum_{(lm)'} \left(W_{+(lm)(lm)'} B_{(lm)'} - i W_{-(lm)(lm)'}
E_{(lm)'}\right),
\end{eqnarray}
where the Hermitian coupling matrices are given by
\begin{eqnarray}
W_{+(lm)(lm)'} &\equiv& \int_S \text{d}S\, W Y^{C*}_{(lm)ab}
Y_{(lm)'}^{C\,ab} \\
W_{-(lm)(lm)'} &\equiv& i\int_S \text{d}S\,W Y^{C*}_{(lm)ab}Y_{(lm)'}^{G\,ab}.
\end{eqnarray}
The matrix $W_{-(lm)(lm)'}$ controls the contamination 
with electric polarization and
can be written as
a line integral around the boundary of the cut defined by $W$. The matrices can be
evaluated easily numerically 
in term of the harmonic coefficients of the window $W_{lm}$
using\footnote{Our sign conventions follow Ref.~\cite{Lewis01}.}
\begin{multline}
W_{\pm (l_1m_1)(l_2m_2)} =  \\
\half(-1)^{m_1}\sum_{l} W_{lm} 
\sqrt{\frac{(2l_1+1)(2l_2+1)(2l+1)}{4\pi}}\times \quad\quad \\
\left[\threej{l}{l_1}{l_2}{0}{2}{-2}\pm
  \threej{l}{l_1}{l_2}{0}{-2}{2}\right] 
\threej{l}{l_1}{l_2}{m}{-m_1}{m_2}
\end{multline}
where $m = m_1 - m_2$.

Vectors are now denoted by bold Roman font,
e.g.\  $\vB$ has components $B_{lm}$, and matrices are denoted by
bold italic font, e.g.\ $\mW_\pm$ have components
$W_{\pm(lm)(lm)'}$. Including the results equivalent to the above for
the E polarization, we then have the harmonic relations
\begin{equation}
\begm \vEt \\ \vBt \enm =  \begm \mW_+ & i\mW_- \\ -i\mW_- &  \mW_+
\enm
\begm \vE \\ \vB \enm.
\end{equation}
Harmonic methods of E/B separation amount to ways of solving these
equations for linear combinations of the observed $\vEt$, $\vBt$ such
as to give results that depend only on $\vE$ or $\vB$. In principle the
matrices are formally infinite, though in practice we measure the components
of $\vEt, \vBt$ up to some finite $\lmax$. However the
pseudo-harmonics to finite $\lmax$ can still can contain
contributions from the underlying fields at all $\ell$ unless the
fields are band limited (i.e. there exists some finite $\lmax$ above
which all the components of $\vE$ and $\vB$ are zero), or the coupling matrices are sufficiently
localized. Thus in general the $\mW_\pm$ coupling matrices are rectangular,
including the coupling to the $E$ and $B$ modes on all scales.

\subsection*{Eigenstructure}

For square coupling matrices the symmetry of the spherical harmonics implies that $\mP \mW_- \mP = - \mW_-^*$ for
the matrix $P_{(lm)(l'm')} \equiv (-1)^m \delta_{ll'}\delta_{-mm'}$ 
which satisfies $\mP^2=1$. This implies that eigenvalues of $\mW_-$
come in $\pm$ pairs:
\begin{equation}
\mW_- \ve = \lambda \ve \quad\implies\quad \mW_- \mP\ve^* = -\lambda \mP\ve^*,
\end{equation}
and hence $\text{Tr}(\mW_-) = 0$. These properties hold for all
$\lmax$ and are a direct result of our window on the sky being a real
function. For a window function which is positive (or zero) 
everywhere $\mW_+$ is positive semidefinite, and the eigenvalues are
bounded between zero and one for window functions normalized to lie between zero and one. 

In the limit that $\lmax \rightarrow \infty$ the coupling matrix
becomes a projection operator onto the section of sky observed 
(from now on we assume the window function is taken to be zero or
one everywhere), so
\begin{equation}
 \begm \mW_+ & i\mW_- \\ -i\mW_- &  \mW_+ \enm =  \begm \mW_+ & i\mW_- \\ -i\mW_- &  \mW_+
\enm^2.
\end{equation}
This property is ensured by the completeness of the harmonics, and
shows that
\begin{equation}
\label{Wids}
\mW_-\mW_+ + \mW_+\mW_- = \mW_- \quad\quad \mW_+^2 + \mW_-^2 = \mW_+.
\end{equation}
For an eigenvector $\ve_\lambda$ of $\mW_+$, with $\mW_+ \ve_\lambda = \lambda
\ve_\lambda$, these relations imply that
\begin{eqnarray}
\mW_-^2 \ve_\lambda &=& \lambda(1-\lambda) \ve_\lambda \nonumber\\
\mW_+[\mW_-\ve_\lambda] &=&
(1-\lambda) [\mW_- \ve_\lambda].
\end{eqnarray}
It follows that $\mW_- \ve_\lambda$ is also an eigenvector of $\mW_+$
with eigenvalue $1-\lambda$, and we can define
$\ve_{1-\lambda}$ so that $\mW_- \ve_\lambda = \sqrt{\lambda(1-\lambda)}
\ve_{1-\lambda}$. Hence the eigenstructure of $\mW_-$ is given by 
\begin{equation}
\label{WMeigs}
\mW_-(\ve_\lambda \pm \ve_{1-\lambda}) = \pm \sqrt{\lambda(1-\lambda)} (\ve_\lambda \pm \ve_{1-\lambda}).\end{equation}

Thus two sets of eigenvectors of $\mW_+$ lie in the null space of $\mW_-$:
\begin{eqnarray}
&\mW_+ \ve_0 = 0 \quad\quad &\mW_- \ve_0 = 0 \\
&\mW_+ \ve_1 = \ve_1 \quad\quad &\mW_- \ve_1 = 0.
\end{eqnarray}
The set of vectors $\ve_1$ form a basis for the set of supported
pure-B and pure-E modes
and the $\ve_0$ form a basis for modes that are zero within
the observed regions and therefore cannot be measured.
The remaining modes with $\lambda \notin \{0,1\}$ form a basis for a
set of ambiguous modes.

As $\{l,l'\} \rightarrow \infty$ the elements of the coupling
matrix $W_{+(lm)(lm)'} \rightarrow W_{(lm)(lm)'}$ where $\mW$ is the
coupling matrix for the scalar harmonics (see Ref~\cite{Mortlock00}). Completeness of the scalar
harmonics implies $\mW$ is a projection
matrix and hence has a fraction $\sim\fsky$ unit eigenvalues and
$\sim (1-\fsky)$ zero eigenvalues, where $\fsky$ is the fraction of
the sky included in the cut. Thus we expect a fraction $\sim
\fsky$ of the 
eigenvalues of $\mW_+$ to be unity and
$\sim (1-\fsky)$ to be zero, with the other eigenvalues making up a
fraction $\propto 1/\lmax$ corresponding to the boundary to area
ratio. 
These results only apply in the limit that $\lmax \rightarrow \infty$,
however for finite $\lmax\agt 100$ we expect a large number of
eigenvalues of $\mW_+$ close to zero or one, corresponding to modes
which are either very well or very poorly supported over the observed area.

\section{Harmonic E/B separation}
\label{separation}
To measure the $B$ only we look for a matrix $\mP_B^\dag$ such that 
\begin{equation}
\mP_B^\dag \begm \mW_+ \\ -i\mW_- \enm = 0. 
\end{equation}
Assuming such a $\mP_B$ can be found, we have then have a vector $\vX_B$ of
pure-B modes
\begin{equation}
\vX_B \equiv \mP_B^\dag \begm \vEt \\ \vBt \enm = \mP_B^\dag \begm i\mW_- \\ \mW_+
\enm \vB
\end{equation}
that have no dependence on $\vE$.  The projection onto $E$ is generated similarly with
\begin{eqnarray}
\mP_E = \Rot \mP_B, 
\end{eqnarray}
where $\mI$ is the identity matrix. 
The pure E and B modes are then given by
\begin{equation}
\begm \vX_E \\ \vX_B \enm = \begm \mP_E^\dag \\ \mP_B^\dag \enm 
\begm \vEt \\ \vBt \enm = \begm \mM \vE \\ \mM\vB \enm
\end{equation}
where
\begin{equation}
\mM = \mP_E^\dag \begm \mW_+ \\ -i\mW_- \enm = \mP_B^\dag \begm i\mW_- \\ \mW_+ \enm.
\end{equation}

\subsection*{Band limited case}
Here we assume there exists some $\lmax$ for which all components of $\vB$, \emph{and
  $\vE$}, are negligible for $\ell>\lmax$. We can perform a singular
  value decomposition (SVD) to write
\begin{eqnarray}
\begm i \mW_- \\ \mW_+ \enm &=& \mU \mD \mV^\dag \approx \mUt \mDt
\mVt^\dag \label{Vdef}\\
\begm \mW_+ \\ -i\mW_- \enm &=& \mR \mU \mD \mV^\dag \approx \mR \mUt \mDt\mVt^\dag
\end{eqnarray}
where
\begin{equation}
\mR \equiv \Rot.
\end{equation}
The matrices $\mU$ and $\mV$ are column unitary, and $\mD$ is
diagonal. Since we have observations over only part of the sky, many
elements of $\mD$ will be close to zero, indicating modes that are not
supported over the observed area. The tilded variables are
constructed by deleting the corresponding columns and rows of $\mU$
and $\mV^\dag$, making $\mDt$ a smaller square matrix. The
approximation can be made the same order as the numerical precision by
choosing the threshold for the diagonal elements of $\mD$ small enough.

Premultiplying by $\mUt^\dag$ we then have
\begin{eqnarray}
\mUt^\dag \Pt &\approx&\mDt\mVt^\dag \vB + i \mH \mDt \mVt^\dag \vE \\
\mUt^\dag \mR^\dag \Pt &\approx&  \mDt\mVt^\dag \vE - i\mH \mDt \mVt^\dag \vB
\end{eqnarray}
where the Hermitian matrix $\mH$ is defined by
\begin{equation}
i\mH \equiv \mUt^\dag \mR \mUt.
\end{equation}
The vector $\mDt\mVt \vB$ contains essentially all the observable
information about $\vB$. Solving we have
\begin{eqnarray}
\mUt^\dag\left[ \mI - \mR \mUt \mUt^\dag \mR^\dag \right] \Pt &\approx&
(1-\mH^2) \mDt \mVt^\dag \vB\\
\mUt^\dag\mR^\dag\left[ \mI -  \mUt \mUt^\dag \right] \Pt &\approx&
(1-\mH^2) \mDt \mVt^\dag \vE,
\end{eqnarray}
which achieves the E/B separation. It is in the form of a projection
operator to remove the left-range of the coupling, followed by a
reduction into the basis of partially supported modes by
multiplication with $\mUt^\dag$.
For band limited skies in the absence of noise there is no
information loss.  However for modes corresponding to non-zero
eigenvalues of $\mH$ the signal to noise is decreased relative to doing no
separation. This is consistent with the understanding that the noise
on the second derivatives of the data needed to perform direct E/B
decomposition becomes larger as you near the boundary of the region. 

Note that this method is not useful for extracting large scale B modes
from CMB polarization observations since even though B may be
effectively band limited at low $\ell$, there is also expected to be a E signal with
a high band limit with $\lmax \agt 10^3$. 
Vectors of harmonics are of size $n = (\lmax+1)^2-4$ which makes the method computationally infeasible for $\lmax
\agt 200$ without simplifying symmetries. On the full sky it is
possible to impose a low band limit by convolution, however on the cut
sky this is not possible without mixing information from inside and outside
the cut.

\subsection*{Non-band limited case}


We now consider what happens to the above results as the band limit is removed.
In the limit $\lmax \rightarrow \infty$ we can use~\eqref{Wids} to show
\begin{eqnarray}
\mH &\approx& -\mDt^{-1} \mVt^\dag \left( \mW_+\mW_- +
   \mW_-\mW_+\right) \mVt \mDt^{-1} \nonumber \\
   &\rightarrow& -\mDt^{-1}\mVt^\dag \mW_- \mVt \mDt^{-1},
\end{eqnarray}
and from~\eqref{Vdef} that $\mVt$ diagonalizes $\mW_+$ with $\mVt^\dag \mW_+ \mVt =
\mDt^2$. Taking $\mVt$ to have the related
eigenvectors $\ve_\lambda$ and $\ve_{1-\lambda}$ in adjacent columns, it also follows from~\eqref{WMeigs} that
$\mVt^\dag \mW_- \mVt$ is block diagonal, where the blocks are either
zero or off-diagonal $2\times 2$ matrices with eigenvalues
$\pm\sqrt{\lambda(1-\lambda)}$. 
This implies that $\mH$ is block
diagonal, with elements {\tiny{$\begm 0 & 1 \\ 1 & 0\enm$}} or zero, and hence that
$\mH^2$ is diagonal, with elements zero or one. Thus each mode in $\mDt\mVt^\dag
\vB$ can either be measured exactly (for the zero eigenvalues of $\mH^2$)  or
is completely lost by the $E/B$ separation (for the unit
eigenvalues). The lost modes that cannot be separated correspond to the
`ambiguous' modes discussed in Ref.~\cite{Bunn03}.
Since the zero eigenvalues of $\mH^2$ are determined by
the null space of $\mW_-$, in this limit
$E/B$ separation amounts to projecting out the non-zero
eigenvalues of $\mW_-$, corresponding to the boundary terms.
For non-band limited skies in the limit $\lmax \rightarrow \infty$ the method for
projecting out the range of $\mW_-$  advocated in Ref.~\cite{Lewis01} 
is therefore optimal for general sky
patches, in addition to being optimal in the simple azimuthal case
analysed in detail (see also Appendix~\ref{azimuthal}). 

\begin{figure}
\begin{center}
\psfig{figure=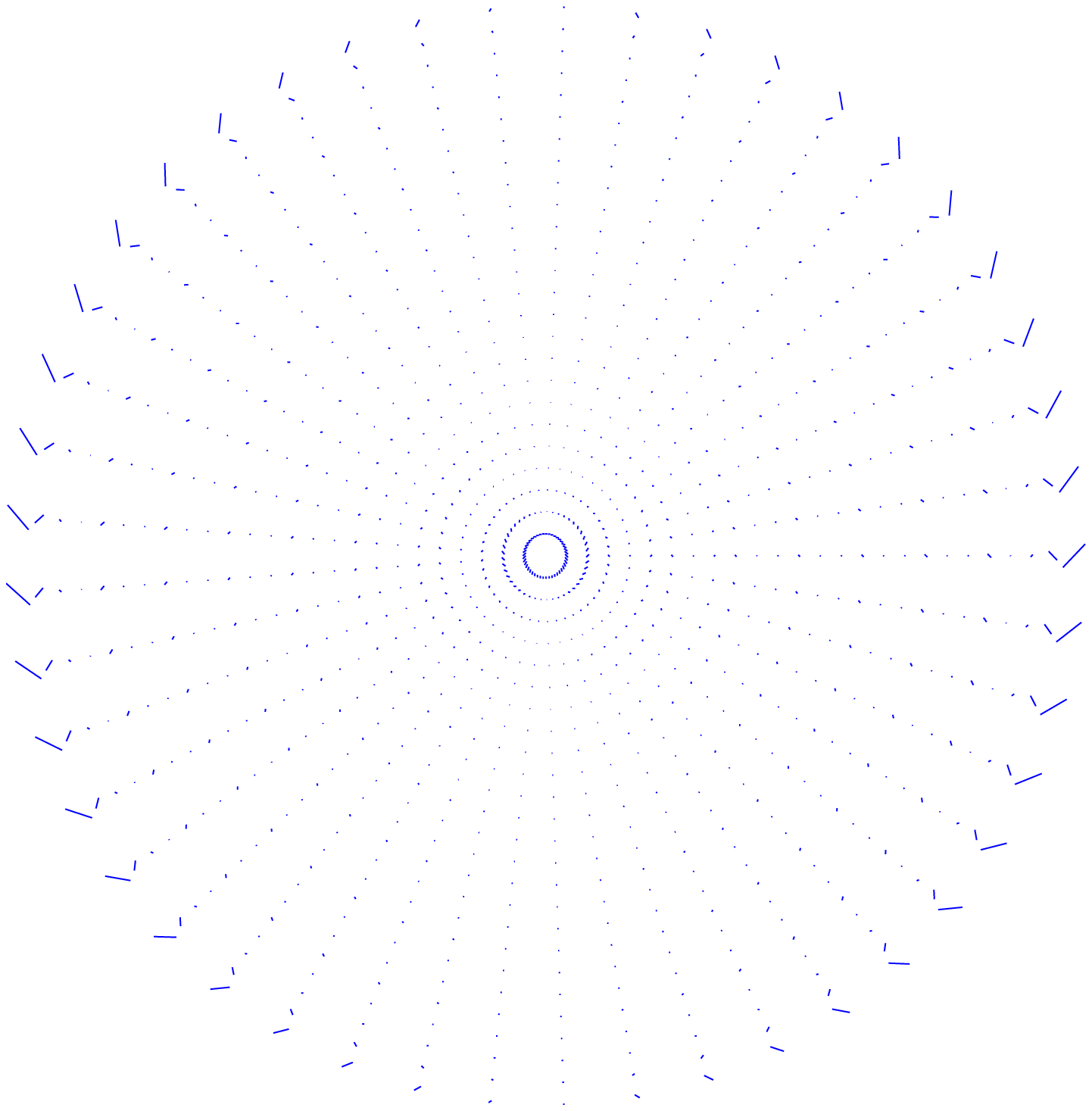,angle=0,width = 5cm}\\
\psfig{figure=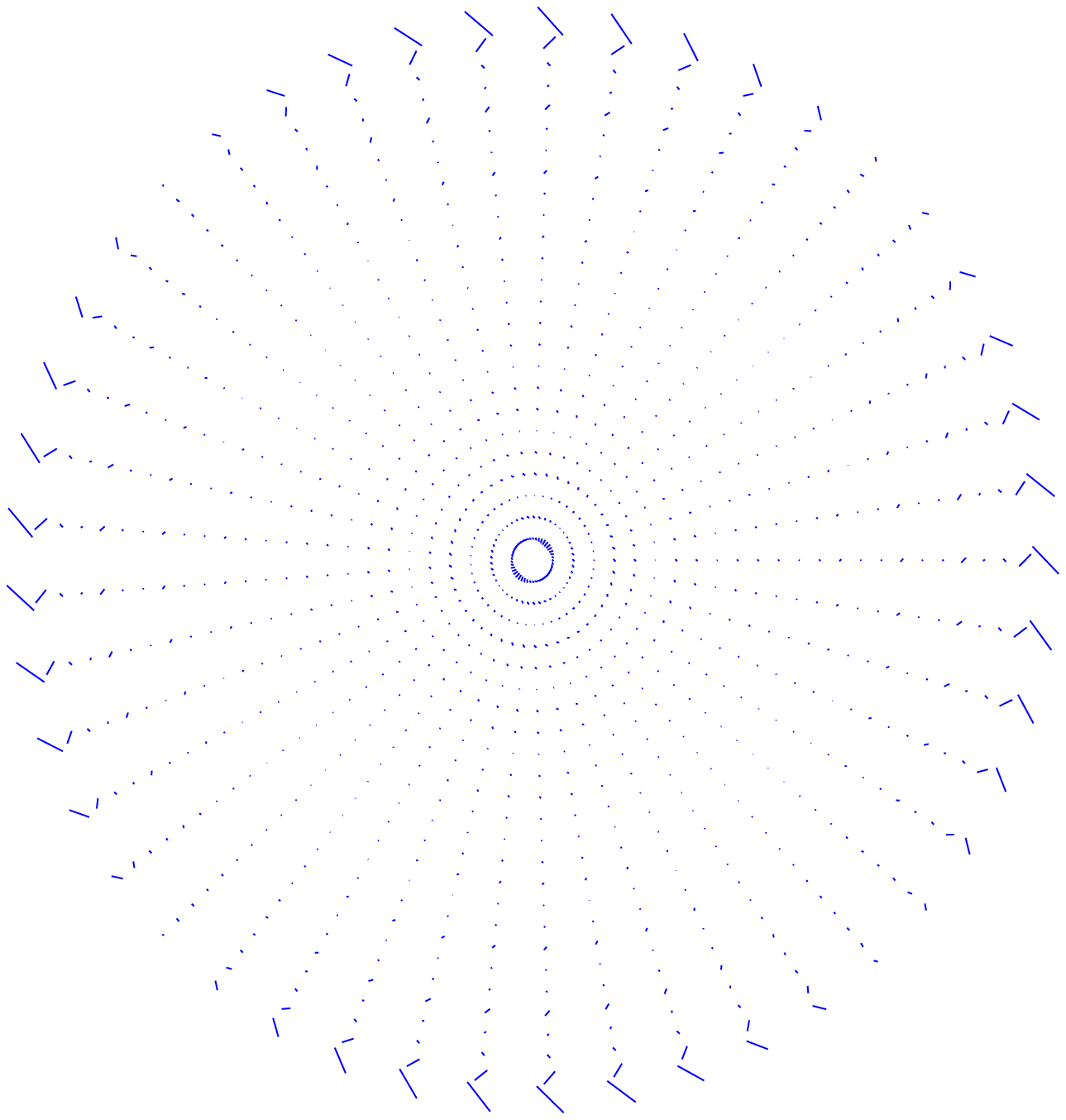,angle=0,width = 5cm}
\caption{
Pure $B$ modes with non-zero
projection into the range of $\mW_-$, for band limited $E$ with
$\lmax=300$. The modes are for an azimuthally symmetric patch of sky
with where $\theta_{\text{max}} =20^\circ$, and have $m=1$ and $m=2$.
\label{newmode}}
\end{center}
\end{figure}

In Fig.~\ref{newmode} we show two modes which for
a band limit of $\lmax =300$ are pure $B$, but have non-zero
projection into the range of $\mW_-$. It is clear that these are
dominated by a line integral around the boundary, and as such they have
significantly worse noise than the other modes due to the
non-zero eigenvalue of $\mH^2$. The line integral is sensitive to E
power on scales with $\ell > \lmax$. As the band limit increases the eigenvalue of
$\mH^2$ tends to one, and the line integrals can no longer be measured
(the signal to noise goes to zero).

For band limited skies one
can use knowledge of the band limit to use the information from the
modes with non-zero eigenvalues of $\mW_-$.
 However in practice the
$E$ polarization is effectively non-band limited as far as extracting
the large scale $B$ signal is concerned. Thus one needs to find modes
which lie in the null-space  $\mW_-$ to extract pure B modes. 
Note
that it is the non-band limitedness of $E$ which implies inevitable
loss due to mode separation, even in the limit of zero noise.

\subsection*{Extracting low multipoles}

\begin{figure}
\begin{center}
\psfrag{l}[][][1.7]{$\ell$}
\psfrag{Cl}[][][1.7]{$C_l / \mu K^2$}
\psfig{figure=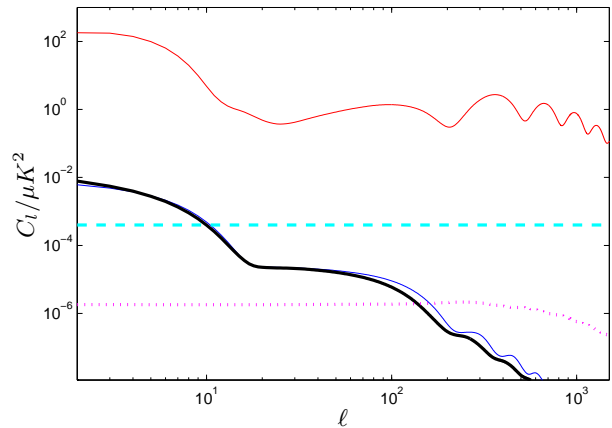,angle=0,width = 8cm}
\caption{Typical $C_l$ for a concordance model with a reionization
  optical depth of $\tau = 0.17$. The top line shows the scalar
  $E$-mode power spectrum, thick and thin solid lines show tensor $B$ and
  $E$ power spectra at an amplitude where tensors contribute about 1/10th
  of the large scale temperature anisotropy ($A_T = 4\times 10^{-10}$). The
  dashed line gives an indication of the noise level with Planck, the
  dotted line show the $B$ spectrum due to gravitational
  lensing of the scalar $E$ signal. The detailed shape of the curves
  on large scales depends on the reionization history.
\label{cls}}
\end{center}
\end{figure}

In the case where the E signal is effectively non-band limited, but we
observe only a finite number $n$ of pseudo-harmonics with $\ell \le
\lmax$, 
a pure $B$ mode $\ve^\dag \vBt$ can
be extracted by finding a vector $\ve$ lying in the left null space of
$\mW_-^\infty$, where $\mW_-^\infty$ is in general an
$n\times\infty$ matrix, coupling in $E$ on all scales. 
This requires that 
\begin{equation}
|\ve^\dag \mW_-^\infty|^2 = \ve^\dag (\mW_+ -\mW_+^\infty\mW_+^\infty{}^\dag ) \ve = 0
\end{equation}
where here the $\mW_+$ is the Hermitian square $n\times n$ matrix, 
$\mW_+^\infty = (\mW_+,\mX)$ and $\mX$ is $n\times \infty$. 
For a supported mode with $\mW_+ \ve = \ve$ this criterion is satisfied
because $|\ve^\dag \mW_+^\infty| \ge |\ve^\dag\mW_+|$ and $|\ve^\dag
\mW_-^\infty|^2$ must be positive or zero.
Note
that $\mW_+\ve = \ve$ is sufficient but not necessary for $|\ve^\dag
\mW_-^\infty|$ to vanish --- in the case of an azimuthal patch there
is a left null space for finite $\lmax$ even though there are no
vectors satisfying $\mW_+ \ve =\ve$ (see
Appendix~\ref{azimuthal}). However in general, without including
information up to the band limit of the data or special symmetries,
this cannot be done. The idea here is to perform E/B separation for
the low multipoles without including data up to the band limit (which
due to the $\lmax^6$ scaling would be infeasible with current
computers for general 
patches). 

In general for finite $n$ there will be no fully supported modes,
however for an eigenvector $\ve$ of $\mW_+$ with  $\mW_+ \ve
= (1 - \epsilon)\ve$ it follows that $|\ve^\dag \mW_-^\infty|^2 = {\cal
  O}(\epsilon)$. In other words, approximately pure $B$ modes
are simply the well supported modes. Note that
$\mW_-\ve \approx 0$ is necessary but not sufficient since it does not ensure
there is no coupling to power from $E$ on scales with $\ell >\lmax$ in
general. 

The signal variance of $\ve^\dag \vBt$ is given by
\begin{equation}
\la |\ve^\dag \vBt|^2\ra = W^+_l C^{BB}_l + W^-_l C^{EE}_l
\end{equation}
where
\begin{equation}
W^+_l \equiv \sum_m | (\ve^\dag \mW_+)_{lm} |^2 \quad\quad 
W^-_l \equiv \sum_m | (\ve^\dag \mW_-)_{lm} |^2.
\end{equation}
The scalar contribution to the $C^{EE}_l$ power spectrum is typically between $1\text{--}100$ times
larger than $C^{BB}_l$ for $\ell \alt 100$ and levels of B detectable by
Planck (see Fig.~\ref{cls}), 
and remains of the same order of magnitude up to $\ell\sim
2000$. Taking $C^{EE}_l \sim \text{const}$ the E contribution to the
variance is $\sim|\ve^\dag \mW_-^\infty|^2$, thus we require $\epsilon
\ll 0.01$ for clean separation of B.
 However there is little point removing E to levels much lower than
 the experimental noise, so for noise limited observations larger
 values of $\epsilon$ could be used.

\begin{figure}
\begin{center}
\psfrag{l}[][][1.7]{$\ell$}
\psfrag{Wl}[][][1.7]{$W_l^\pm$}
\psfig{figure=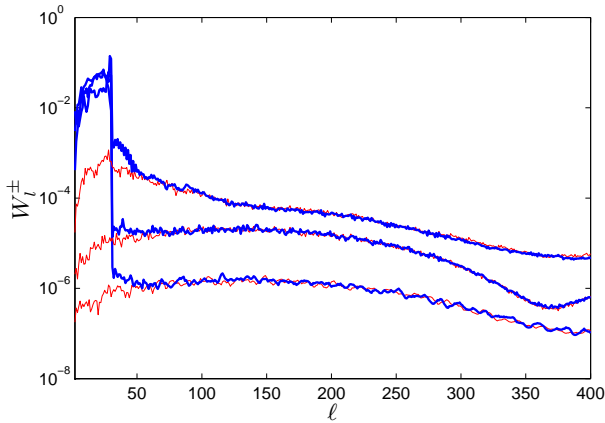,angle=0,width = 8cm}
\caption{The functions $W_l^+$ (thick lines, the coupling to B) and
  $W_l^-$ (thin lines, the coupling to E) giving the contributions to the
  variance for some nearly pure B modes
  with various degrees of support ($\epsilon = \{0.1, 0.01, 0.001\}$,
  top to bottom), constructed with $\lmax = 30$ and the asymmetric sky
  cut discussed in the text.
\label{wins}}
\end{center}
\end{figure}

In Fig.~\ref{wins} we show a few window functions $W^+_l$ and $W^-_l$ for nearly pure B
modes, using $\lmax = 30$ and the realistic cut discussed in the next
section. This low value of $\lmax =30$ is not sufficient to extract
many well supported modes (there is only one with $\epsilon = 0.001$),
however it 
allows us to compute the rectangular matrices $\mW_+$ and $\mW_-$ up
to much 
higher $\ell$ in order to explicitly show the coupling of higher multipoles as a
function of the choice of $\epsilon$. The
expected behavior is demonstrated in the figure, with small values of $\epsilon$
effectively removing the coupling to E on all scales.

\section{Examples}

\begin{figure*}
\begin{center}
\psfig{figure=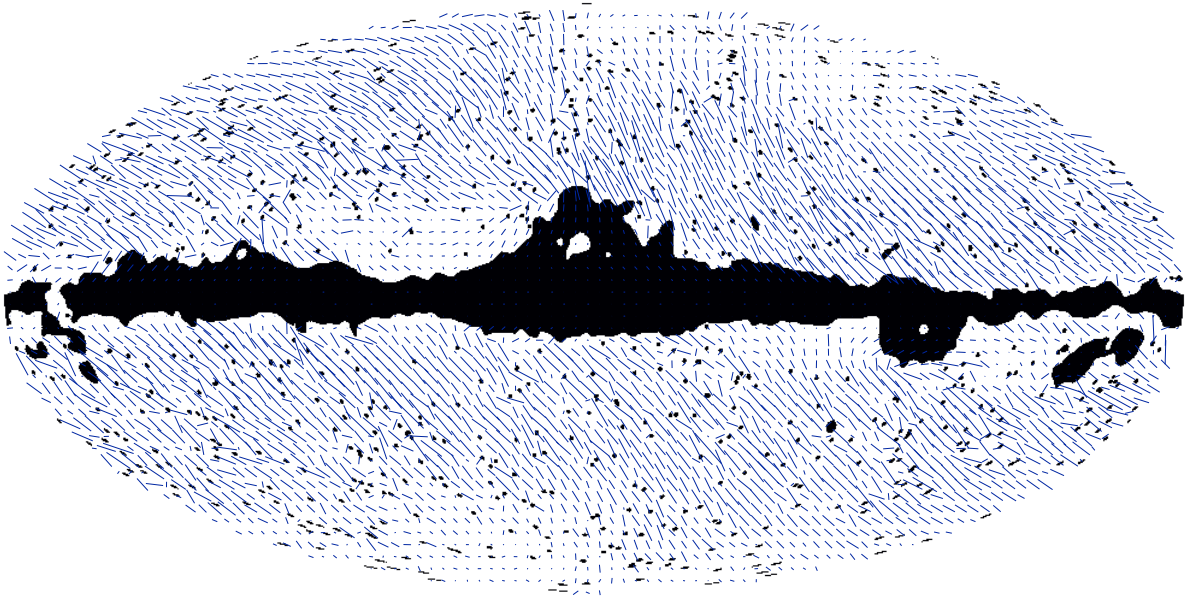,angle=90,width = 16cm,clip=}
\caption{The nearly pure well supported B mode with maximal signal for
  a $\tau\sim 0.12$ reionization model, $\lmax=150$, $\epsilon=0.001$ and the `kp2' sky cut discussed in the text. The Mollweide projection is used for the cut, but polarization orientations are not projected.
\label{polmap}}
\end{center}
\end{figure*}

We now demonstrate explicitly the extraction of large scale B modes from a
realistic non-symmetric sky cut, the `kp2' 
cut\footnote{\url{http://lambda.gsfc.nasa.gov/}} used
by the WMAP analysis~\cite{Bennett03}. This excludes the galactic region, in addition
to a number of other foreground sources, in total excluding about
$15\%$ of the sky, and is shown as the solid area in Fig.~\ref{polmap}. This may not be a good cut for polarized
observations, but the cut is realistic in the sense that it does not
have any artificial symmetries, and therefore is a reasonable test
case for how B mode extraction can work in practice. 

As shown in Fig.~\ref{cls} the large scale
reionization signal with $\ell \alt 10$ is large and has high signal
to noise, at least if the optical depth really is $\agt 0.1$.
We use $\lmax =150$, 
which is computationally tractable ($\sim 10\text{GB}$ of memory, few
days of CPU time) and sufficient to get most of the large scale
reionization power. The well supported mode with $\epsilon = 0.001$ and maximal signal is shown in Fig.~\ref{polmap}. 
With more effort (e.g. distributing the
computation over a cluster or more efficient algorithms) it may be possible to push to
$\lmax \sim 300$, possibly enough to extract essentially all of
the separable B modes due to primordial gravitational
waves.  This may be essential if the reionization turns out to be less
significant, and of course resolving the shape of the magnetic power
spectrum is a key check that the signal really is due to primordial
tensor modes. 
Separation on smaller scales (e.g. as a consistency
check, and for studying the lensing signal) would have to proceed on
azimuthally symmetric cuts over clean regions of the sky with a number
of cuts around foreground sources (see Appendix~\ref{azimuthal}), or 
rely on quadratic methods that work very well for estimating the separated power
spectra on small scales. 

Identifying the well supported modes is straightforward, and
requires diagonalization of the Hermitian matrix $\mW_+$, 
\begin{equation}
\mW_+ = \mU_+ \mD_+ \mU_+^\dag.
\end{equation}
Since we only need the well supported modes, in practice we only need to
compute the eigenvectors
with eigenvalues\footnote{\samepage{We use the LAPACK routine `ZHEEVR', see \url{http://www.netlib.org/lapack}}} between $1-\epsilon$ and $1$,
for some choice of $\epsilon$ and $\lmax$. 
For large $\lmax$ there are
$\sim\fsky\lmax^2$ such modes, where $\fsky$ is the fraction of the
sky included in the window. For less large $\lmax$ a significant
fraction of modes will only be partially supported; losing the subset
of these modes that are not contaminated with $E$ is the price we pay
for a computationally tractable analysis with manageable
$\lmax$. The well supported eigenvectors define a reduced column-orthogonal matrix $\mUt_+$ which projects the cut-sky harmonics into the space of well supported nearly-pure B modes:
\begin{equation}
\vX_B\equiv \mUt_+^\dag \vBt \approx \mUt_+^\dag \vB.
\end{equation}
A map of the nearly pure B modes can then be constructed using the harmonics $\vB' = \mUt_+ \mUt_+^\dag \vBt$. Note that though computing $\mUt_+$ is time consuming for large $\lmax$, once this has been done for a particular cut, construction of B maps for different realizations is computationally fast. 

\begin{figure*}
\begin{center}
\psfig{figure=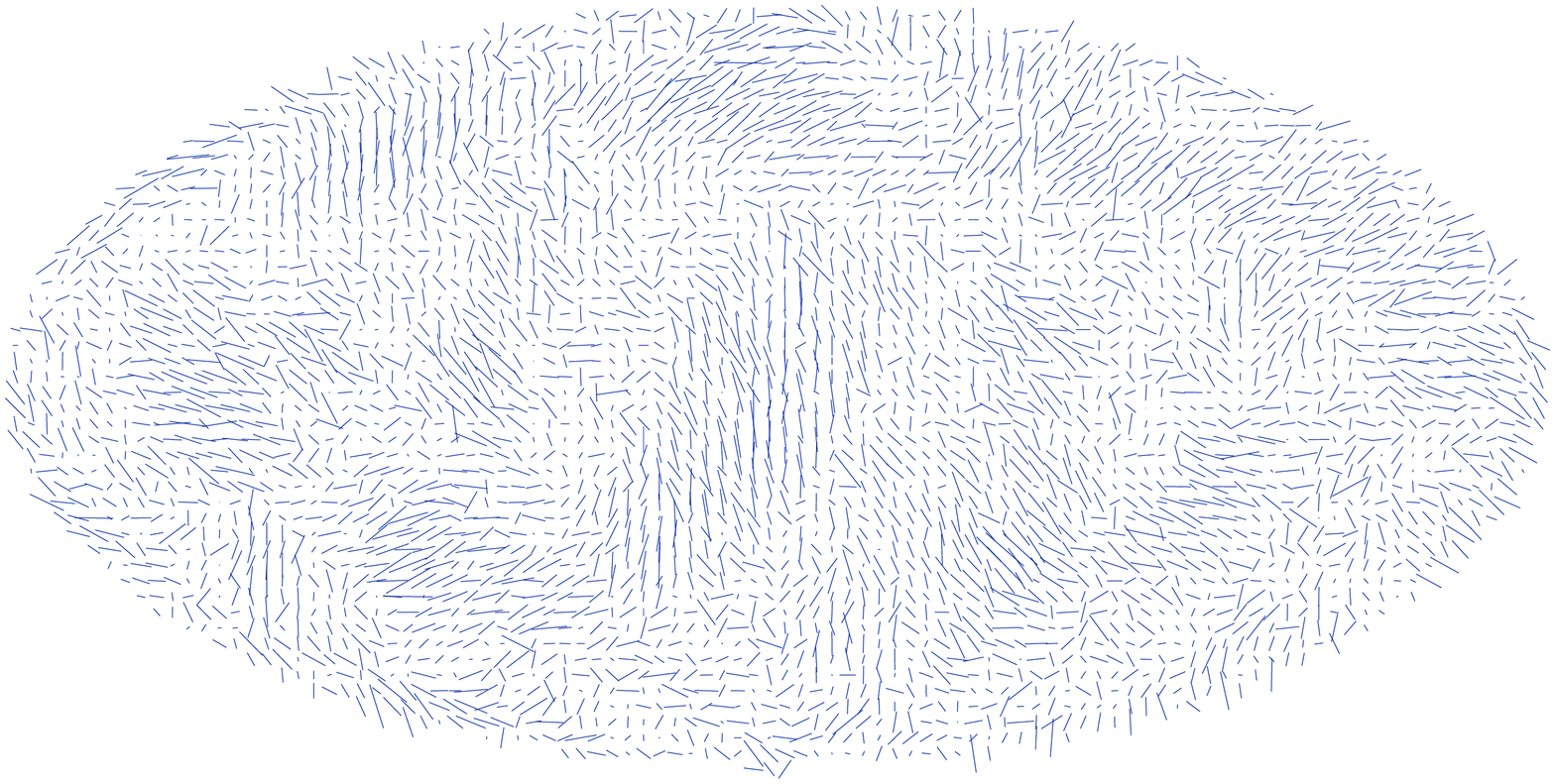,angle=90,width = 13cm,clip=}
\psfig{figure=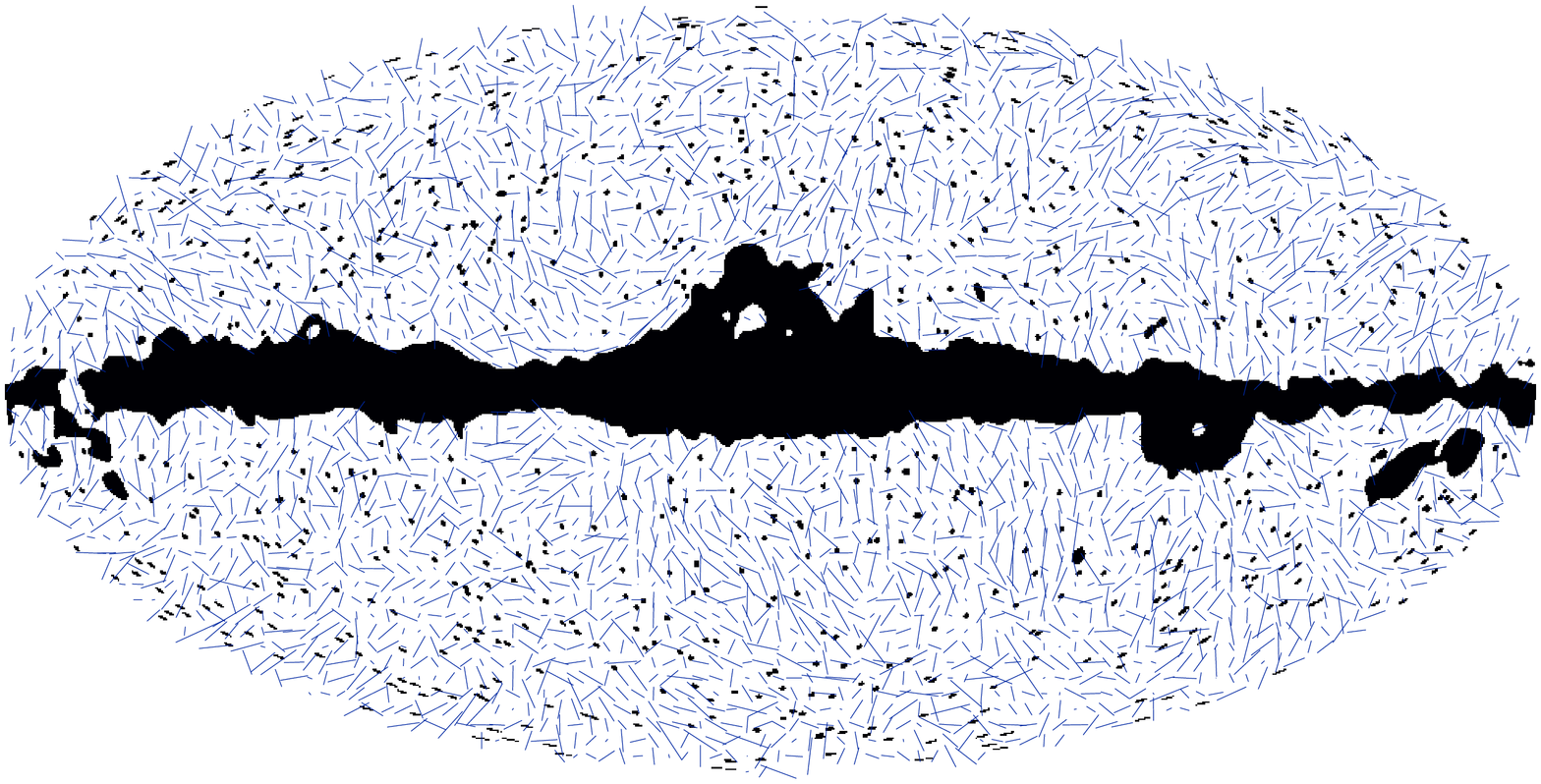,angle=90,width = 13cm,clip=}
\psfig{figure=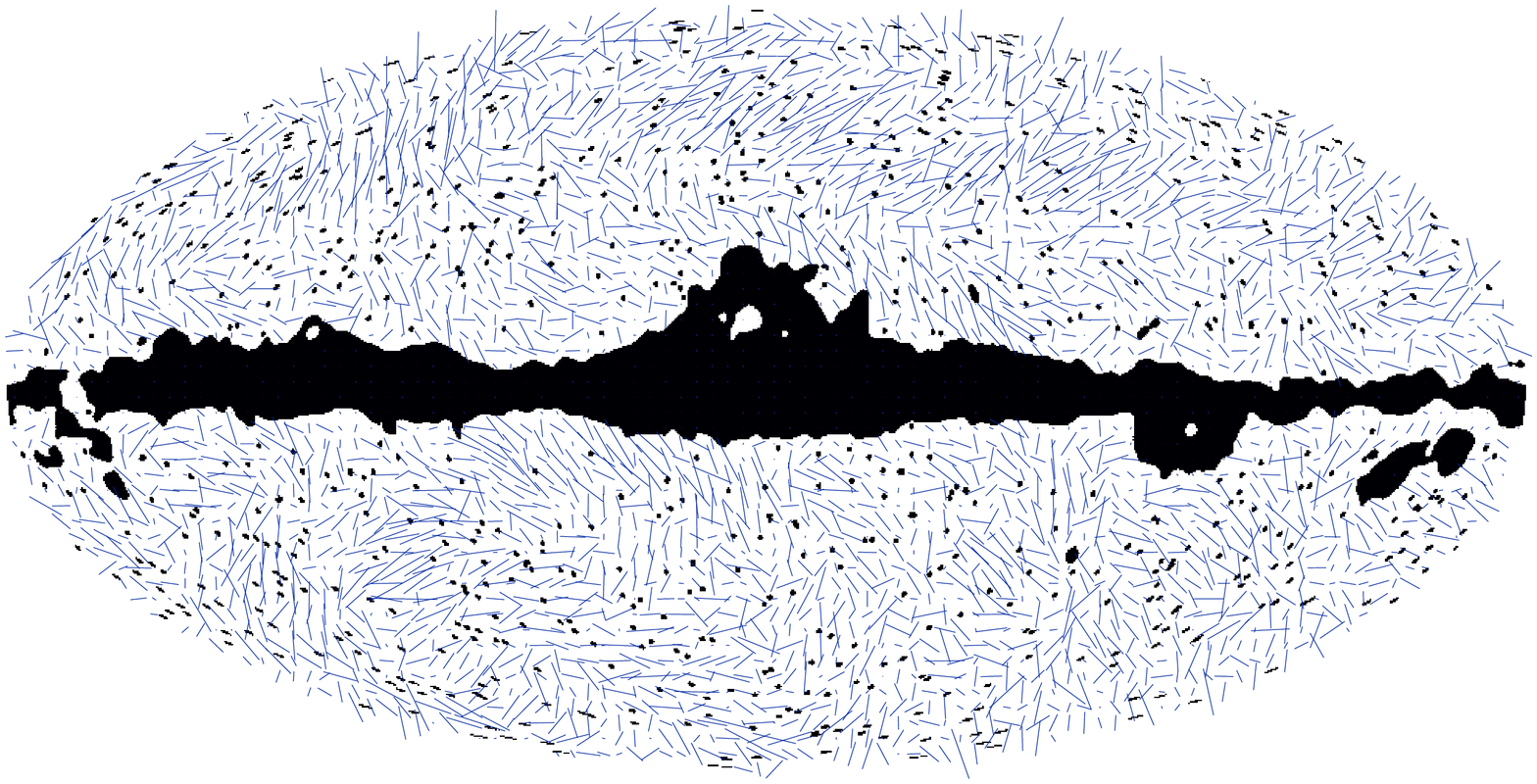,angle=90,width = 13cm,clip=}

\caption{Simulation of noise-free well supported B mode extraction ($\lmax=150$, $\epsilon=0.001$) for the `kp2' cut. Top: input B map due to tensor modes. Middle: the observed cut sky including E polarization from scalar and tensor modes ($A_T = 4\times 10^{-10}$). Bottom: the reconstructed well supported nearly pure B modes. Line orientations show the polarization direction, line lengths show the magnitude. For each map the lengths are independently normalized. The sky is represented in the Mollweide projection (but polarization orientations are not projected), and the beam size is half a degree.
\label{maps}}

\end{center}
\end{figure*}

In Fig.~\ref{maps} we show the construction of a B map for a simple simulation without noise\footnote{Further color maps are available at \url{http://cosmologist.info/polar/EBsupport.html}}. The recovered B map is not the same as the input map since we are only using $\lmax=150$, and some modes are lost due to E/B separation. 
However it is clear than many of the large scale features are present in the recovered map, and that the method works usefully well.

\subsection*{Noise}

In general there will be anisotropic and correlated noise on the polarization measurements. The E/B separation methods we discussed give E/B clean separation of the signal, however any interpretation of the separated modes would have to carefully model the actual noise properties in the observation under consideration. In general the E and B modes will have complicated noise correlations. However in the idealized case of isotropic uncorrelated noise on the Stokes' parameters the noise properties are straightforward~\cite{Lewis01}. In particular the well supported E and B modes will also have uncorrelated isotropic noise
\begin{eqnarray}
\la \vX_B \vX_B^\dag \ra_N  &\approx &\la \mUt_+^\dag \vB \vB^\dag \mUt_+ \ra_N \approx \la \vX_E \vX_E^\dag\ra_N \propto \mI \nonumber \\
\la \vX_B \vX_E^\dag \ra_N  &\approx& \la \mUt_+^\dag \vB \vE^\dag \mUt_+ \ra_N \approx 0.
\end{eqnarray}
The complications introduced by correlated anisotropic noise do not effect the \emph{signal} E/B separation, which is purely determined by the geometry of the sky cut. As the noise in a region inside the cut increases, some of the modes will become much more noisy.  In the limit of infinite noise in this region, corresponding to an additional cut, these modes have zero signal to noise and can be removed. This removes combinations of E and B modes in such a way that the number of pure B modes that can be measured decreases, and new ambiguous modes (which are combinations of E and B with finite noise) are generated.

\subsection*{Detection by Planck?}
\label{Planckex}
\begin{figure}
\begin{center}
\psfrag{Tensor Amplitude AT}[][][1.5]{Tensor Amplitude $A_T$}
\psfrag{Detection Probability}[][][1.5]{Detection Probability}
\psfig{figure=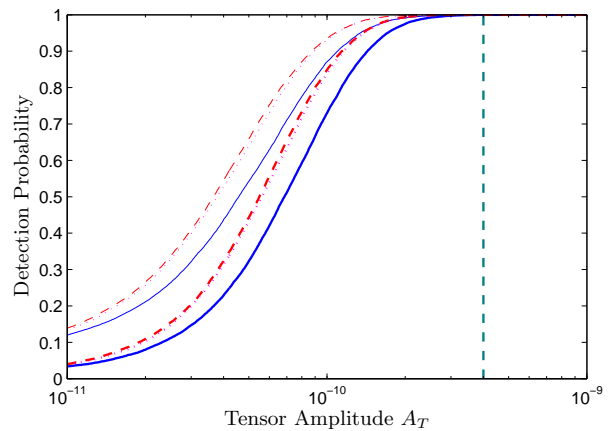,angle=0,width = 8cm}
\caption{Tensor magnetic polarization detection probabilities for
  Planck at a confidence of 99 per cent (thick lines) and 95
  per cent (thin lines) using nearly pure B modes. Solid lines are for the `kp2' cut using
  $\lmax =150$ and $\epsilon = 0.001$. Dashed lines are for exact E/B
  separation with an
  azimuthally symmetric cut (with the same $\fsky$) and $\lmax = 250$. Dotted
  lines are for the azimuthally symmetric cut, but using the well
  supported modes method ($\lmax =150$ and $\epsilon = 0.001$).
 For all lines we assume isotropic $C_{\text{noise}} = 4.1\times
  10^{-4}\mu K^2$ and a reionization optical depth $\tau \sim 0.12$.
   The vertical dashed line shows the tensor
  amplitude that would contribute about 1/10th of the large scale
  temperature anisotropy (approximately the limit that can be set
  without using polarization). 
\label{Planck}}
\end{center}
\end{figure}

Fig.~\ref{Planck} shows how the E/B separation method performs with regards to
the detection probability\footnote{We follow the method used in
  Ref.~\cite{Lewis01}. Due to the high signal to noise of a few of the
large scale modes this may be somewhat suboptimal, and therefore pessimistic.} for magnetic polarization due to tensor
modes, taking a simple model of the Planck satellite as a test case
(assuming isotropic Gaussian noise, that foregrounds can be subtracted
accurately outside the cut region, and systematics are negligible). We
assume the primordial tensor modes are Gaussian with scale invariant
power spectrum, with $A_T$ being the variance of the
transverse traceless part of the metric tensor, and that other
cosmological parameters are well known.

The approximate method applied to the asymmetric cut
does not perform quite as well as an optimal analysis with an
azimuthally symmetric cut, but
the difference is not that large --- equivalent to reducing a  $99\%$ confidence
detection to about $95\%$. The asymmetric cut is not expected to
perform as well because it
increases the area adjacent to a boundary and hence the number of ambiguous
modes (though the approximate result may be improved
slightly by taking larger $\lmax$). The plot also shows the result of
applying the non-exact method (retaining just the well supported
modes) to the azimuthally symmetric cut, and shows that the results
are almost identical to performing the exact separation in this case.
In the exact
azimuthal analysis the result is almost completely insensitive to
$\lmax$, with $\lmax=30$ results lying on top of those shown. This is
because of the small number of large signal to noise modes on large
scales for a significant optical depth to reionization. In the asymmetric case much larger $\lmax$ is
needed to obtain well supported modes.

\begin{figure}
\begin{center}
\psfrag{Tensor Amplitude AT}[][][1.5]{Tensor Amplitude $A_T$}
\psfrag{Detection Probability}[][][1.5]{Detection Probability}
\psfig{figure=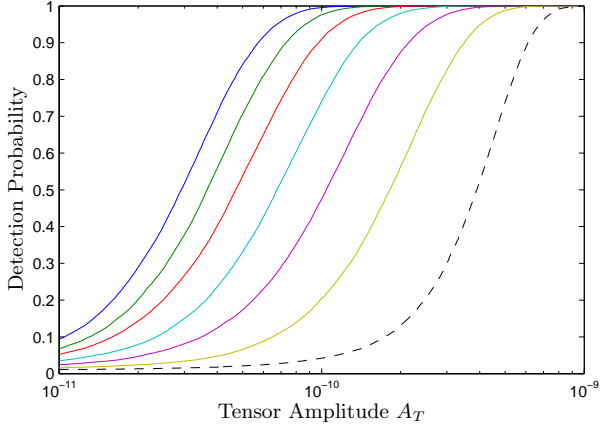,angle=0,width = 8cm}
\caption{The probabilities for 99 per cent confidence detections of
  magnetic polarization with a simple model of Planck,  for
  reionization optical depths $\tau = \{
  0.2,0.17,0.14,0.11,0.08,0.05\}$  (left to right) assuming sharp
  reionization. The dotted line shows the result if there is no
  foreground-distinguishable B mode signal at $l <30$.
Exact B mode separation
  is used with an azimuthally symmetric galactic cut with $\fsky = 0.85$.
\label{zre}}
\end{center}
\end{figure}

For $\tau \agt 0.1$ it is clear that Planck should have a good chance
of detecting tensor amplitudes $A_T \agt 10^{-10}$, corresponding to
an energy scale $V^{1/4} \agt 10^{16} \text{GeV}$ at horizon crossing
during inflation. The sensitivity to tensor magnetic modes however depends on the
reionization optical depth, as shown in Fig.~\ref{zre}. 
If the optical depth turns out to be at the lower end, the
asymmetric cut with $\lmax \sim 150$ would not perform so well
compared to the exact azimuthal case since there would be significant
power on smaller scales, requiring $\lmax \gg 100$ to construct all the
well supported modes.

\section{Conclusion}

We have demonstrated explicitly that E/B separation is possible on
large scales for realistic non-symmetric sky cuts. If the reionization
optical depth is large almost all the detection significance for
tensor modes would come from the very largest scales, where mode
mixing is most important. We showed that these modes can be separated
in a computationally tractable way
by retaining only the well supported modes. The ambiguous modes that
cannot be separated will be dominated by the E signal if the noise is low, and
whilst useful for analysis of the E polarization would not add
significantly to a full likelihood analysis for magnetic
polarization.

With map-level separation one can also in principle study the full
distribution of the B signal, for example testing for
Gaussianity. Analysing a pure B map is essentially the same as
analysing a temperature map, and much more more straightforward and
conceptually cleaner than doing a full joint analysis.  
For noise limited observations, when the loss of ambiguous modes loses a
significant amount of information,  an analysis with E/B
separated variables still provides a valuable cross-check on results from a 
full joint analysis, and may help to isolate systematics. 

\begin{acknowledgments}
I thank Ue-Li Pen for stimulating discussions, Anthony Challinor for
pointing out an important error in an early draft, and Sarah Bridle
for helpful suggestions. I acknowledge use of the
HEALPix\footnote{\url{http://www.eso.org/science/healpix/}}~\cite{Gorski:1998vw}
and LAPACK packages.
\end{acknowledgments}

\appendix

\section{Circular boundaries}
\label{azimuthal}
For a circular boundary at constant latitude $\theta=\Theta$
(i.e.\ the boundary of an azimuthal patch), modes with different $m$
decouple, $\mW_-$ is block diagonal, and modes with different $m$ can be handled separately.
The $\mW_-$ surface boundary integral evaluates to~\cite{Lewis01}
\begin{equation}
W_{-(lm)(l'm')} = -\frac{m\delta_{mm'}}{2|m|} \left[u_l(m) u_{l'}^\ast(m) + v_l(m) v_{l'}^\ast(m)\right]_\Theta,
\label{eq:wminusdcmp}
\end{equation}
where the vectors
\begin{eqnarray}
u_l(m) &\equiv& \sqrt{\frac{(l-2)!}{(l+2)!}}\sqrt{8|m|\pi}
\sin\theta\frac{\d}{\d\theta}\left( \frac{Y_{lm}}{\sin\theta}\right)
\\
v_l(m) &\equiv& \sqrt{\frac{(l-2)!}{(l+2)!}}\frac{\sqrt{8|m|\pi(m^2-1)}}{\sin\theta} Y_{lm}
\end{eqnarray}
for $l\geq 2$ and some arbitrary $\phi$. In the
$\lmax\rightarrow\infty$ limit this is the spectral decomposition
($v_l(m)$ and $u_l(m)$ are normalized and orthogonal), which follows from
 the results:
\begin{eqnarray}
\sum_{l=|m|}^\infty \frac{(l-2)!}{(l+2)!} | Y_{lm}|^2 =
\frac{ \sin^2\theta}{8|m|\pi(m^2-1)} \\ 
\sum_{l=|m|}^\infty \frac{(l-2)!}{(l+2)!} Y_{lm}^*
\sin\theta\frac{\d}{\d\theta}\left(\frac{Y_{lm}}{\sin\theta}\right)
= 0 
\end{eqnarray}
for $|m| \ge 2$, and 
\begin{eqnarray}
\sum_{l=\max(2,|m|)}^\infty
\frac{(l-2)!}{(l+2)!}\left|\sin\theta\frac{\d}{\d\theta}\left(\frac{Y_{lm}}{\sin\theta}\right)
\right|^2 = \frac{1}{8|m|\pi}
\end{eqnarray}
for $|m| \ge 1$.
In this limit the eigenvalues are $-\half m/|m|$, a special case of the
general analysis given in section~\ref{harmonics}. For a single
circular boundary and finite $\lmax$
there are at most two non-zero eigenvalues per $m$, with one for
$|m|=1$ and none for $m=0$. 

The azimuthal case has the nice property that even if $\vEt$ and
$\vBt$ are only available for finite $\lmax$ one can project out
exactly the cross-contamination on all scales. The matrix
$\mW_-$ truncated at a finite number of rows (but retaining an
unlimited number of columns) can be written as
\begin{equation}
\mW_- = \mU_- \mD_- \mV_-^\dag
\end{equation}
where $\mU_-$ is a column unitary matrix with at most two columns per $m$, and hence
the range of $\mW_-$ can be projected out using
\begin{eqnarray}
\mP_- &\equiv \mI-\mU_-\mU_-^\dag 
\end{eqnarray}
so that
\begin{eqnarray}
\mP_- \vEt &= \mP_- \mW_+ \vE \quad\quad\quad \mP_- \vBt &= \mP_- \mW_+ \vB
\end{eqnarray}
losing at most two modes per $m$ (per boundary). The matrix $\mU_-$
can in practice be constructed for each $m$ by normalization and
orthogonalization (e.g. using a SVD) of the two 
vectors $u_l(m)$ and $v_l(m)$ (which are no longer orthogonal or
normalized for finite $\lmax$). 
This projection is equivalent to the procedure of projecting out the
non-zero eigenvalues of $\mW_-$ in Ref.~\cite{Lewis01}, for $|m| \le
\lmax - 2n$, where $n$ is the number of boundaries.
If $\mW_-$ is
smaller than it's range due to a finite $\lmax$ (i.e. for
$|m|>\lmax-2n$)  the zero eigenvalues $\mW_-$ do not correspond to
separated modes.

Two maps consisting of pure
$E$ and pure $B$ can be constructed simply from
\begin{eqnarray}
\vEt'  &= (\mI-\mU_-\mU_-^\dag) \vEt,  \quad\ \vBt' =& 0\quad
(\text{pure E})\quad\\
\vBt' &= (\mI-\mU_-\mU_-^\dag)\vBt, \quad \vEt' =& 0 \quad (\text{pure B}).\quad
\end{eqnarray}
If desired a map of the ambiguous modes can also be constructed from
the remaining modes. The separation is computationally trivial in the
azimuthal case due to the separability in $m$, and is possible up to
the resolution of the experiment (typically $\lmax > 10^3$).

For $n$ co-axial circular boundaries one
looses up to $4n(\lmax-n)$ 
modes due to the E/B separation, and the separation remains exact and
separable in $m$.
For non-co-axial boundaries exact separation is still possible,
even though modes with different $m$ are
mixed due to the rotation (i.e. with the Wigner-D matrices
$D^l_{mm'}$). This is because the coupling
matrix $\mW_-$ for the entire sky made up of a set of non-intersecting
cuts can be written as
 a finite sum of
coupling matrices $\mW_-$ each with finite range, and therefore itself has finite range
which can be projected out exactly. 
Thus it is still possible to
perform exact E/B separation at finite $\lmax$ with azimuthal cuts containing a finite
number of non-intersecting circular cuts around foreground
sources. 

To extract a set of E and B modes for likelihood evaluation one can
proceed following Ref.~\cite{Lewis01}. We briefly review the method
here with one minor enhancement. First we diagonalize $\mW_+$ (taking
it to be square) and
discard the badly supported modes, to write $\mW_+ \approx \mUt_+
\mDt_+ \mUt_+ ^\dag$ so
\begin{eqnarray}
 \mDt^{-1/2}_+ \mUt_+^\dag\vEt  \approx \mDt^{1/2}_+ \mUt_+^\dag \vE + i
 \mDt^{-1/2}_+ \mUt_+^\dag \mW_- \vB \\
 \mDt^{-1/2}_+ \mUt_+^\dag\vBt  \approx \mDt^{1/2}_+ \mUt_+^\dag \vB - i
 \mDt^{-1/2}_+ \mUt_+^\dag \mW_- \vE.
\end{eqnarray}
Then we do the diagonalization
\begin{equation}
 \mDt^{-1/2}_+ \mUt_+^\dag \mW_- \mUt_+ \mDt^{-1/2}_+ = \mU_- \mD_- \mU_-^\dag
\end{equation}
and construct the matrix $\mUt_-$ by deleting the columns of $\mU_-$
corresponding to non-zero diagonal elements of $\mD_-$. It 
follows from the Hermiticity of $\mW_-$ that
$\mUt_-$ is made up of vectors in the left null-space of the mode-mixing
matrix $\mDt^{-1/2}_+ \mUt_+^\dag \mW_-$.  
The pure E and B modes are then given by
\begin{eqnarray}
\mUt_-^\dag \mDt^{-1/2}_+ \mUt_+^\dag\vEt  &\approx& \mUt_-^\dag\mDt^{1/2}_+
\mUt_+^\dag \vE \\
\mUt_-^\dag \mDt^{-1/2}_+ \mUt_+^\dag \vBt  &\approx& \mUt_-^\dag\mDt^{1/2}_+
\mUt_+^\dag \vB.
\end{eqnarray}
For isotropic uncorrelated noise these modes also have isotropic and
uncorrelated noise. This method is equivalent to that
presented in Ref.~\cite{Lewis01} but slightly faster as it replaces
the SVD of the $ \mDt^{-1/2}_+ \mUt_+^\dag \mW_-$ coupling matrix
with a faster diagonalization of a smaller Hermitian matrix by using
the Hermiticity of $\mW_-$. Note that as mentioned above one should
only include modes with $|m| \le \lmax -2n$, though in practice for
galactic cuts these modes are badly supported anyway. The coupling
matrices for an azimuthal cut can be computed efficiently using the
results given in the appendix of Ref.~\cite{Lewis01}.

\vfill



\end{document}